\documentclass[prl,twocolumn,superscriptaddress]{revtex4}

\usepackage{graphicx}
\usepackage{latexsym}
\usepackage{bm}
\usepackage{amsmath}

\newcommand{\comment}[1]{}

\newcommand{\etal}{{\it et al.}}
\begin{document}

%\draft

\title{Elliptic flow of colored glass in high energy heavy ion collisions} 

\author{Alex Krasnitz}
\affiliation{FCT and CENTRA, Universidade do Algarve,Campus de Gambelas,
   P-8000 Faro, Portugal}
\author{Yasushi Nara}
\affiliation{%
 RIKEN BNL Research Center, Brookhaven National Laboratory,
                Upton, N.Y. 11973, U.S.A.
}
\author{Raju Venugopalan}
\affiliation{%
 c\ Physics Department, Brookhaven National Laboratory, Upton, N.Y. 11973,
U.S.A.
}
\affiliation{%
 RIKEN BNL Research Center, Brookhaven National Laboratory,
                Upton, N.Y. 11973, U.S.A.
}

\date{\today}

\begin{abstract}
We compute the elliptic flow generated by classical gluon fields in 
a high energy nuclear collision. The classical gluon fields are described by 
a typical momentum scale, the saturation scale $\Lambda_{s}$, which is, for RHIC energies, 
of the order of $1-2$ GeV. A significant elliptic flow is generated
only over time scales on the order of the system size $R$.
The flow is dominated by soft modes $p_{\text{T}}\sim  \Lambda_{s}/4$
 which linearize at 
very late times $\tau\sim R \gg 1/\Lambda_{s}$. 
We discuss the implications of our result for the theoretical interpretation of the 
RHIC data.
\end{abstract}

\pacs{25.75.-q, 24.10.-i, 24.85.+p}

\maketitle

\section{Introduction}

The collective flow of excited nuclear matter has been an important 
tool in attempts to extract the nuclear equation of state ever since the early
days of heavy ion collision experiments~\cite{eos}. 
Measurements of collective flow at the Relativistic Heavy Ion Collider (RHIC) 
may provide insight into the excited partonic matter, often called a 
Quark Gluon Plasma (QGP), produced in high energy heavy ion 
collisions~\cite{QMproc}. 
In particular, the azimuthal anisotropy in the transverse momentum distribution
has been proposed as a sensitive probe of the hot and dense matter
produced in ultra-relativistic heavy ion collisions~\cite{Ollitraut}. 
A measure of the azimuthal anisotropy is the second Fourier coefficient of 
the azimuthal distribution, the elliptic flow parameter $v_2$. 
Its definition is~\cite{Voloshin} 
\begin{eqnarray}
v_2 = \langle \cos(2\phi) \rangle = 
\frac{ \int^{\pi}_{-\pi} d\phi \cos(2\phi) \int d^2p_{\text{T}}
                \frac{d^3N}{dy d^2p_{\text{T}}d\phi}}
       {\int^{\pi}_{-\pi} d\phi  \int d^2p_{\text{T}}
                \frac{d^3N}{dy d^2p_{\text{T}}d\phi}  }.
\label{eqn1}
\end{eqnarray}
The elliptic flow for non-central collisions is believed to be sensitive to the
early evolution of the system~\cite{Sorge}.

The first measurements of elliptic flow from RHIC, at center of mass
energy $\sqrt{s_{NN}}=130$ GeV, have been reported
recently~\cite{STARv2}.  Hydrodynamic model calculations provide good
agreement, for large centralities, and for particular initial
conditions and equations of state, with the measured centrality
dependence of the data~\cite{KHTS}.
The agreement at smaller centralities is less good, 
perhaps reflecting the breakdown of a hydrodynamic description 
in smaller systems~\cite{Hirano:2001eu}.
Hydrodynamic models also agree well with the 
$p_{\text{T}}$ dependence of the unintegrated (see Eq.~(\ref{eqn1})) 
elliptic flow parameter $v_2(p_{\text{T}})$ up to 1.5 GeV/c at 
mid-rapidity~\cite{KHTS}. However, 
above $1.5$ GeV, the experimental distribution 
appears to saturate, while the hydrodynamic model distributions continue 
to rise~\cite{KHTS}.
Jet quenching scenarios to 
explain this saturated behavior of $v_2 (p_{\text{T}})$ at large
$p_{\text{T}}$ ~\cite{v2jet} appear to disagree quantitatively with the data.
Partonic  transport models including only elastic
gluon-gluon scattering require large cross sections or 
large initial gluon number to obtain significant  
elliptic flow~\cite{Molnar:2001ux}.

In this letter, we compute the contribution to $v_2$ at central
rapidities from the strong color fields generated in the initial
instants after the heavy ion collision. These are generated as
follows.  At high energies, or equivalently, at small Broken $x$, the
parton density in a nucleus grows very rapidly and
saturates eventually~\cite{GLRMQ} forming a Color Glass Condensate
~\cite{MV} (CGC).  The CGC is characterized by the color charge 
squared per unit area $\Lambda_s^2$ which grows with 
energy, centrality and the size of the nuclei.
Estimates for RHIC give $\Lambda_s\sim 1-2$ GeV~\cite{comment0}.  For a
recent review of the CGC model and additional references, 
see Ref.~\cite{McLerran:Review}.
Since the occupation number of gluons in the CGC is large, $f\sim
1/\alpha_S(\Lambda_s^2)>1$, classical methods can be applied to study
gluon production in heavy ion collisions at high
energies~\cite{KMW,KN} . The energy
and number~\cite{AR99,AR01} of gluons produced were computed
numerically for an SU(2) Yang--Mills gauge theory and recently
extended to the SU(3) case~\cite{AYR01}. We have confirmed that 
strong electric and magnetic fields of order
$1/\alpha_S$ are generated in a time $\tau \sim 1/\Lambda_s$ 
after the collision.

The classical Yang--Mills approach may be applied to compute elliptic
flow in a nuclear collision~\cite{comment1}. For peripheral nuclear collisions, the 
interaction region is a two-dimensional almond shaped region, with the $x$ axis lying 
along the impact parameter axis and the $y$ direction perpendicular to it and to the 
beam direction. We will show that even though 
large electric and magnetic fields (and the corresponding transverse components of the 
pressure in the x and y directions) are generated over very short time scales $\tau\sim 1/\Lambda_s$, 
the significant differences in the pressures, responsible for elliptic flow, 
are only built up over much longer time scales $\tau \sim R$. Moreover, the elliptic flow 
is generated by soft modes $p_{\text{T}}\sim \Lambda_s/4$. Our result has important consequences for 
the theoretical interpretation of the RHIC data-these will be discussed later in the text.

\section{Numerical Method}

We now discuss our numerical computation of elliptic flow. As in our earlier work, 
we assume strict boost invariance. The dynamics is then that of a Yang-Mills gauge field
coupled to an adjoint scalar field in 2+1-dimensions. 
For a numerical solution we use lattice discretization.
The discretized theory is described by a Kogut-Susskind 
Hamiltonian~\cite{AR99}.

In previous work, we studied gluon production in central collisions of
very large nuclei and therefore assumed a uniform color charge 
distribution ($\Lambda_s=constant$) in the transverse plane. To study effects of anisotropy and
spatial inhomogeneity, we shall consider a finite nucleus. We shall impose  
suitable neutrality conditions on the distribution of color
sources~\cite{Lam:1999} to prevent gluon production at large distances 
outside the nucleus.

To this end, we model a nucleus as a sphere of radius $R$, 
filled with randomly distributed nucleons of radius 
$l\approx 1\,$fm. For a gold nucleus, $R\approx 6.5\,$fm. 
The color charge distribution within a nucleon is generated as follows.
First, we generate (throughout the transverse plane of a nucleon) a random
uncorrelated Gaussian distribution $\rho^a(\vec r)$ ($a$ being the adjoint 
color index and $\vec r$ the transverse-plane position vector), obeying
the relation 
$$\langle\rho^a(\vec r)\rho^b(\vec r')\rangle=\Lambda_{\rm n}^2\delta^{ab}
\delta(\vec r-\vec r')$$ 
where the $\langle\rangle$ average is over the ensemble of nucleons.
Next, we remove the monopole and dipole components of the distribution by
superimposing the distribution with the appropriate homogeneous contribution; 
first of the color charge, then
of the color dipole moment. For a sufficiently fine lattice discretization, 
this procedure does not result in a significant change in the average magnitude
of the random charge distribution. Since the color charges of different
nucleons are uncorrelated, the resulting nuclear color charge squared 
per unit area has a position-dependent magnitude,
$$\Lambda_s^2(r)=\frac{2}{l}\Lambda_{\rm n}^2\sqrt{R^2-r^2}$$, 
where $r$ is the transverse radial coordinate relative to the beam axis through
the center of the nucleus and $l$ is the nucleon diameter.
We can adjust $\Lambda_{\rm n}$ to ensure the central nuclear color 
charge squared per unit area $\Lambda_{s0}^2\equiv \Lambda_s^2(0)$ has a 
desired value.

Once the color charge distributions of the incoming nuclei are 
determined, the corresponding classical gauge fields can be computed. 
The initial conditions for the gauge fields in the 
overlap region between the nuclei are obtained as discussed 
previously~\cite{AR99}.
For each configuration of color charges sampled, 
Hamilton's equations are solved on the lattice for 
the gauge fields and their canonical momenta as a function of the 
proper time $\tau$.

\section{Results}

\begin{figure}
\includegraphics[width=3.0in]{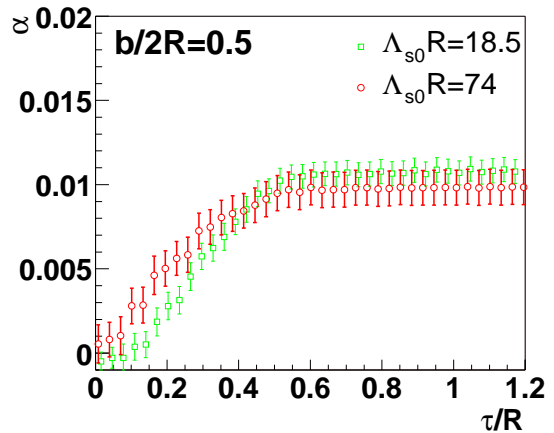}
\caption{The momentum anisotropy parameter 
$\alpha = \langle T^{xx}-T^{yy}\rangle/\langle T^{xx}+T^{yy}\rangle$
for a peripheral nuclear collision corresponding to impact parameter 
$b/2R =0.5$ is 
plotted versus the proper time $\tau$ in units of the nuclear radius $R$ for 
two different values of $\Lambda_{s0}R$.
}
\label{fig:time}
\end{figure}

We first compute the momentum anisotropy parameter $\alpha$ 
(defined in Fig.~\ref{fig:time})
as a function of the proper time $\tau$~\cite{comment3}. 
The results for values of the 
external parameter $\Lambda_{s0}R=18.5$ and $\Lambda_{s0}R=74$ (spanning 
the RHIC-LHC range of energies) are shown in Fig.~\ref{fig:time}. We observe 
that $\alpha$ rises gradually saturating at $\alpha\sim 1$\% at a proper time 
on the order of the size of the system. The time required to develop an 
anisotropy is clearly much larger than the characteristic time 
$\sim 1/\Lambda_{s0}$ associated with non-linearities in the system. 

The calculation of elliptic flow, defined by Eq.~(\ref{eqn1}), involves 
determining the gluon number, a quantity whose meaning 
is ambiguous outside a free theory.
Closely following our earlier work~\cite{AR01,AYR01}, we resolve this 
ambiguity by computing the number in two different ways; 
directly in Coulomb Gauge (CG)
and by solving a system of relaxation (cooling) equations for the fields.
Both definitions give the usual particle number in the case of a free theory.
We expect the two to be in good agreement for a weakly coupled theory.
Wherever the two disagree strongly, we should not trust either.
%%%%%%%%%%%%%%%%%%%%%%%%%%%%%%%%%%%%%%%%%%%%%%%%%%%%%%%%%%%%%%%%%%%%%%%%%%%
%%%%%%%%%%%%%%%%%%%%%%%%%%%%%%%%%%%%%%%%%%%%%%%%%%%%%%%%%%%%%%%%%%%%%%%%%%

It is easy to show that $v_2\,N$, $N$ being the total gluon number, can
be reconstructed from the cooling time history of $T_{xx}-T_{yy}$, just as $N$
can be reconstructed from that of the energy functional~\cite{AR01}:
\begin{equation}
   v_2\, N = \sqrt{\frac{2}{\pi}}\,
\int^{\infty}_0 \frac{dt}{\sqrt{t}} (T^{xx}(t) - T^{yy}(t)) \, . 
\label{v2cool}
\end{equation}
This expression for $v_2\, N$ is manifestly gauge invariant.

In contradistinction to the gluon number, an estimate of $v_2$ 
involves both the fields and their conjugate momenta. 
Indeed, consider the expression for $T_{xx}-T_{yy}$ in our system:
$$T_{xx}-T_{yy}=\int{\rm d}^2x_\perp\left[E_y^2-E_x^2+(D_x\Phi)^2-(D_y\Phi)^2
\right],$$
where $\bm{E}$ is the chromo-electric field,
 $\Phi$ the adjoint scalar field, and
$D_i$ the covariant derivative, In the weak-coupling limit $D_i$ reduces
to $\partial_i$, the usual derivative. In that limit, the first two terms in
$T_{xx}-T_{yy}$ only involve the conjugate momenta of gluons polarized in the
transverse plane, while the last two terms only depend on the fields of gluons
whose polarization is perpendicular to that plane. Since it is not a priori 
obvious that the two polarizations contribute equally to $v_2$, both the fields
and the conjugate momenta should be computed. For the 
cooling method, relaxation equations for conjugate momenta  
require that the usual relation between the momenta and 
{\it proper} time derivatives of the fields hold at 
all {\it cooling} times~\cite{AK97}.

\begin{figure} %[b!]%Fig. 2.
\includegraphics[width=3.0in]{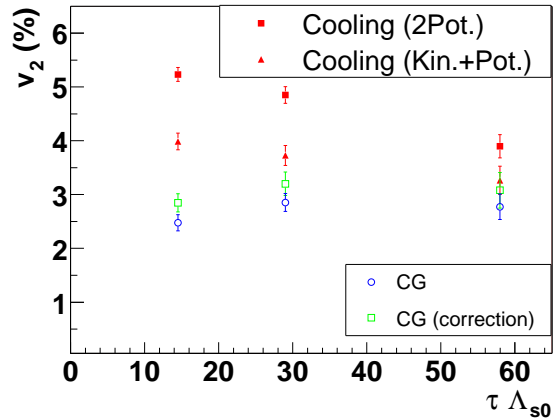}
\caption{$v_2$ is plotted (for impact parameter $b/2R =0.5$) 
versus $\Lambda_s\tau$. Solid squares are cooling 
results using only the potential contribution. Solid diamonds include 
the contribution of both potential and kinetic terms. The Coulomb gauge 
result, including both the potential and the kinetic contributions,
is shown in open circles. Open squares are corrected CG results-see 
text for an explanation.} 
\label{fig:Systematics}
\end{figure}

In Figure~\ref{fig:Systematics}, we compare, for a fixed impact
parameter ($b/2R=0.5$), the values of $v_2$ obtained by the different methods.
In the cooling approach, $v_2$ can be computed
first by considering only the potential part of $T_{xx}-T_{yy}$ in
Eq.~\ref{v2cool} and then assuming 
an equal contribution from the kinetic part. As seen in 
Figure~\ref{fig:Systematics}, such an equality does not hold
until very late times. There is a significant difference at early times
between the CG and cooling estimates of $v_2$.

At asymptotically large cooling times we expect $N$ and
$v_2N$ of the cooled configuration to vanish. If the CG
values of these do not vanish, then they are artifacts of the CG. 
We subtract the residual values from the corresponding values before 
cooling. The result is referred to as the corrected CG values.
The cooling and the CG computations are expected to agree at
late times, as the system becomes increasingly weakly
coupled. The two methods agree for $N$ at fairly early times. For $v_2$, this
convergence occurs at much later times, since, as we shall
see in the following, $v_2$ is dominated by very soft modes
with momenta $p_{\rm T}<\Lambda_{s0}$.

\begin{figure} %[b!]%Fig. 3.
\includegraphics[width=3.0in]{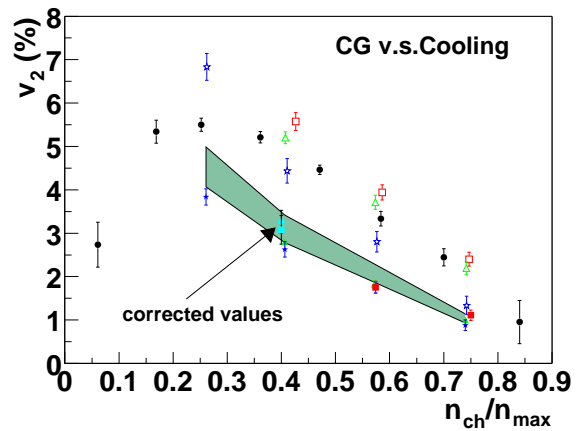}
\caption{The centrality dependence of 
$v_2$ at the earliest times in Fig.~\ref{fig:Systematics} 
is computed using cooling (open symbols) and CG 
(filled symbols). Results are for 
$\Lambda_{s0}R$ spanning the RHIC-LHC range, specifically, 
$\Lambda_{s0} R=18.5$ (squares), 37 (triangles), and 74 (stars). 
Full circles denote {\it preliminary} STAR data~\cite{Snellings}. 
The band denotes the estimated value of $v_2$ when extrapolated to 
very late times. ``Corrected values'' denotes the late time cooling and CG 
result for $\Lambda_{s0}R=18.5$ at one centrality value. 
}
\label{fig:v2CentDep}
\end{figure}

In Figure~\ref{fig:v2CentDep} we plot $v_2$ reconstructed from the cooling
time history of only the potential terms in 
$T_{xx}-T_{yy}$, along with the CG values (also including potential terms only)
as a function of $n_{\text{ch}}/n_{\text{tot}}$ for different 
values of $\Lambda_{s0} R$ as discussed in the figure. The systematic 
errors represented by the band (for $\Lambda_{s0}=18.5$ ~\cite{comment4}) 
are primarily due to limited resources 
available to study the slow convergence of the cooling and CG  
computations. We have studied the late time behavior of $v_2$
for one impact parameter-the results are shown in the figure.

The asymptotic values of $v_2$, as predicted by the model, undershoot the
data.  This disagreement notwithstanding, our results show that a
significant $v_2$ can be generated by the classical fields. For very peripheral collisions,
where the gluon density may be too low to justify the classical
approximation, the predictions of the model are not reliable. 
Interestingly, the dependence of $v_2$ on $\Lambda_{s0} R$ is rather weak. For a fixed
impact parameter, the model predicts that, as
$\Lambda_{s0}R\rightarrow \infty$, the classical
contribution to the elliptic flow goes to zero. This is because 
increasing $\Lambda_{s0} R$ is equivalent to increasing
$R$ for fixed $\Lambda_{s0}$ and therefore reducing the initial anisotropy.

\begin{figure} %[b!]%Fig. 3.
\includegraphics[width=3.0in]{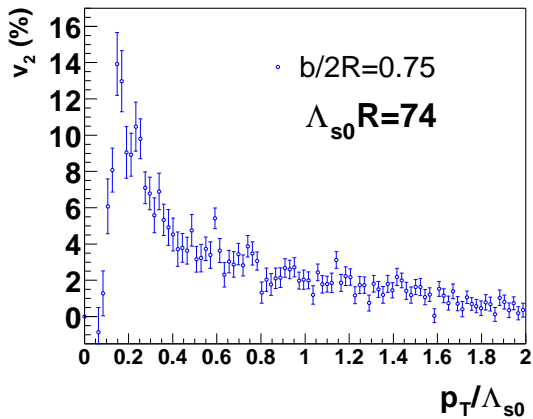}
\caption{$v_2(p_{\text{T}})$ as a function of transverse momentum
 in dimensionless units for $\Lambda_{s0}R=74$.
}
\label{fig:v2dndpt512}
\end{figure}

In Fig~\ref{fig:v2dndpt512}, $v_2 (p_{\text{T}})$ is plotted for
$b/2R=0.75$ for $\Lambda_{s0}R=74$.  Our calculations show that the
elliptic flow rises rapidly and is peaked for $p_{\text{T}}\sim
\Lambda_{s0}/4$ before falling rapidly.  The theoretical
prediction~\cite{TV} is that for $p_{\text{T}} \gg \Lambda_{s0}$,
$v_2(p_{\text{T}})\sim \Lambda_{s0}^2/p_{\text{T}}^2$.  The lattice numerical data
appear to confirm this result-better statistics are required to
determine the large momentum behavior accurately.  
A couple of comments about our result are in order.
Firstly, even though $\Lambda_{s0}^2$ is large, it may differ
considerably from the color charge squared in the region where the
nuclei overlap. This may explain in part why the momenta are peaked at smaller
values of $p_{\text{T}}$. Secondly, the dominant contribution of very
soft modes to $v_2$ helps explain why the cooling and CG 
computations differ until very late times. The soft gluon modes have
large magnitudes and therefore continue to interact strongly until
very late proper times. Concomitantly, the occupation number of these modes 
is not small and the classical approach may be adequate to describe these 
modes even at the late times considered. 

\section{Discussion}

We now turn to the theoretical interpretation of the RHIC $v_2$ data in the 
CGC approach. It is clear from Fig.~\ref{fig:v2CentDep} that our result for $v_2$ contributes 
only about 50$\%$ of the measured $v_2$ for various centralities. 
Our $p_{\text{T}}$ distributions also clearly disagree with experiment~\cite{STARv2,Voloshin:2002ii}.
Naively, one could argue that the classical Yang-Mills approach is only applicable at early times 
so additional contributions to $v_2$ will arise from later stages of the collision. While there is 
merit in this statement, it is also problematic as we will discuss below.

The reason the situation is complex is as follows. We observed that it takes a long time $\tau\sim R$ 
to obtain a significant elliptic flow. At these late times, one would expect that the classical approach 
would be inapplicable due to the rapid expansion of the system. On the other hand, we have seen that 
$v_2$ in the classical approach is dominated by soft modes which are strongly interacting and don't 
linearize even at time scales $\tau\sim R$. Clearly, the soft modes cannot be treated as on-shell partons 
even at times $\tau\sim R$! This is the message one obtains from Fig~\ref{fig:Systematics}.

The correct way to treat the theoretical problem may be as follows. Hard modes with $k_t\geq \Lambda_{s}$ 
linearize on very short time scales $\tau\sim 1/Q_s$. Their subsequent evolution is treated incorrectly in 
the classical approach, which has them free streaming in the transverse plane. In actuality, they are 
scattering off each other via elastic $gg\rightarrow gg$
 and inelastic $gg\leftrightarrow ggg$ collisions 
which drive them towards an isotropic distribution~\cite{BMSS}. This dynamics would indeed provide an 
additional {\it pre-equilibrium} contribution to $v_2$ and is calculable. An effect to consider here would 
be the possible screening of infrared divergences in the hard scattering by the time dependent classical field. 
More complicated is the effect of these hard modes on the classical dynamics of the soft modes and on 
their possible modification of the contribution of the latter to $v_2$. One has here a little explored 
dynamical analog to the interplay of hard particle and soft classical modes in the 
kinetic theory of Hard Thermal Loops~\cite{JPED}.

Were the system to thermalize, both these effects would complement the hydrodynamic component of 
elliptic flow~\cite{KHTS}. A quantitative study of the anisotropy generated in 
this intermediate regime would therefore be very useful in our theoretical understanding of the 
data. 

Finally, we note that $v_2$ is extracted only indirectly   
from a variety of techniques-in particular, two and four particle cumulant
analyses~\cite{Nicola}. Recently, it has been proposed that non-flow 
two particle correlations explain the $v_2$ data~\cite{KT}. 
It is unclear whether this model can explain
other features of the measured azimuthal anisotropy. In our approach,
a procedure very similar to the experimental approach can be followed
and two and four particle correlations can be determined. 
This work will be reported in the near future.

We thank K. Filiminov, U. Heinz, D. Kharzeev, R. Lacey, Z. Lin, 
L. McLerran, A. Mueller, J.-Y. Ollitrault, K. Rajagopal, J. Rak, 
S. Voloshin, Nu Xu, E. Shuryak and D. Teaney for useful comments.
We also thank Sourendu Gupta for contributing to the early stages of
this work. R. V.'s research is supported by DOE Contract
No. DE-AC02-98CH10886. A.K. and 
R.V. acknowledge support from the Portuguese FCT, under grants
CERN/P/FIS/40108/2000 and CERN/FIS/43717/2001.
R.V. and Y.N. acknowledge RIKEN-BNL for support.
A.K. thanks the BNL NTG  for hospitality.
We also acknowledge support in part of NSF Grant No. PHY99-07949.


\begin{thebibliography}{99}

\bibitem{eos} G. F. Bertsch and S. Das Gupta,
    Phys. Rep. {\bf 150}, 189 (1988).

\bibitem{QMproc}
   For reviews and recent developments, see
   {\it Quark Matter '99} [Nucl. Phys. {\bf A661} (1999)];
   {\it Quark Matter 2001},
%    Proceedings of the 15th International Conference
%       on Ultra-Relativistic Nucleus-Nucleus Collisions,
%       Stony Brook, USA, January, 2001,
       [Nucl. Phys. {\bf A698}, (2002)].

\bibitem{Ollitraut} J.-Y. Ollitraut, 
         Phys. Rev. {\bf D46}, 229 (1992).
%	 Phys. Rev. {\bf D48},1131 (1993).

\bibitem{Voloshin}
 S. Voloshin and Y. Zhang, Z. Phys. {\bf C70}, 665 (1996);
 A.M. Poskanzer and S. Voloshin, Phys. Rev. {\bf C58}, 1671 (1998).

\bibitem{Sorge} H. Sorge,
     Phys. Rev. Lett. {\bf 78}, 2309 (1997);
     Phys. Rev. Lett. {\bf 82}, 2048 (1999).

\bibitem{STARv2} STAR Collaboration, K.H. Ackermann \etal,
  Phys. Rev. Lett. {\bf 86}, 402, (2001).

\bibitem{Snellings}We thank R. J. Snellings for providing us with the 
{\it preliminary} STAR data on centrality dependence of $v_2$.

%% hydro %%%%%%%%%%%%%%%%%%%%%%%%%%%%%%%%%%%%%%%%%%%%%%%%%%%%%%%%%%%%%%%%%%

\bibitem{KHTS} P.F.Kolb, P. Huovinen, U. Heinz,and H. Heiselberg, Phys.\
 Lett.\  {\bf B500}, 232  (2001); 
D.~Teaney, J.~Lauret and E.~V.~Shuryak,
Phys.\ Rev.\ Lett.\  {\bf 86}, 4783 (2001); 
arXiv:nucl-th/0110037; {\it ibid.}, Nucl.\ Phys.\ A {\bf 698}, 479 (2002); 
P. F. Kolb, J. Sollfrank, and U. Heinz, Phys.  \ Rev. {\bf C62}, 054909 (2000).

\bibitem{Hirano:2001eu}
T.~Hirano,
%``Is early thermalization achieved only near midrapidity at RHIC?,''
Phys.\ Rev.\ C {\bf 65}, 011901 (2002).
%[arXiv:nucl-th/0108004].

%% Jet quenching %%%%%%%%%%%%%%%%%%%%%%%%%%%%%%%%%%%%%%%%%%%%%%%%%

\bibitem{v2jet}
%HIGH P(T) AZIMUTHAL ASYMMETRY IN NONCENTRAL A+A AT RHIC.
M. Gyulassy, I. Vitev, X.-N. Wang, Phys. Rev. Lett. {\bf 86}, 2537 (2001); 
X.-N. Wang, Phys. Rev. {\bf C63}, 054902 (2001). 

%% transport apprach for v2 %%%%%%%%%%%%%%%%%%%%%%%%%%%%%%%%%%%%%%%%%%%%%

%\bibitem{Zhang:1999rs}
%B.~Zhang, M.~Gyulassy and C.~M.~Ko,
%%``Elliptic flow from a parton cascade,''
%Phys.\ Lett.\ B {\bf 455}, 45 (1999).

\bibitem{Molnar:2001ux}
D.~Molnar and M.~Gyulassy,
%``Saturation of elliptic flow at RHIC: Results from the covariant elastic  parton cascade model MPC,''
Nucl.\ Phys.\ A {\bf 697}, 495 (2002).

\bibitem{GLRMQ}
 L.V. Gribov, E. M. Levin and M. G. Ryskin, Phys. Repts. {\bf 100} (1983) 1;
 A. H. Mueller and J.-W. Qiu, Nucl. Phys. {\bf B268}(1986) 427;
 J. P. Blaizot and A. H. Mueller, Nucl. Phys. {\bf B289} (1987) 847.

\bibitem{MV}
L. McLerran and R. Venugopalan,
Phys.\ Rev.\ {\bf D49} 2233 (1994); {\bf D49} 3352 (1994); {\bf D50} 2225 
(1994); {\bf D59} 094002 (1999); 
J. Jalilian--Marian, A. Kovner, L. McLerran and H. Weigert,  
Phys. Rev. {\bf D55} 5414 (1997);
Y.~V.~Kovchegov, Phys.\ Rev.\ D {\bf 54}, 5463 (1996).  
J. Jalilian-Marian, A. Kovner, and H. Weigert, 
Phys. Rev. {\bf D59} 014015 (1999);
E.~Iancu and L.~D.~McLerran,
Phys.\ Lett.\ B {\bf 510}, 133 (2001).

\bibitem{McLerran:Review}
E.~Iancu, A.~Leonidov and L.~McLerran,
%``The colour glass condensate: An introduction,''
arXiv:hep-ph/0202270.

\bibitem{comment0}The correct leading logarithmic relation between 
$\Lambda_s$ and the gluon saturation scale $Q_s$ is  
$Q_s^2 = \Lambda_s^2 N_c \ln(\Lambda_s^2/\Lambda_{QCD}^2)/4\pi$. 
Assuming (for cylindrical nuclei!)
that $\Lambda_s$ has the same $x$ and atomic number dependence as $Q_s$, 
the Golec-Biernat-Wusthoff parametrization of 
HERA data~\cite{Golec} gives $\Lambda_s\sim 1.4$ GeV. 
Other estimates give $\Lambda_s\sim 2$ GeV~\cite{ActPol}. For realistic 
nuclei, the average value of $\Lambda_s$ over the nucleus may be lower, about $1$ GeV for 
a central value of $1.4$ GeV.


\bibitem{Golec}K. Golec-Biernat and M. Wusthoff, Phys.\ Rev.\ D {\bf 59} (1999) 
014017; A. Stasto, K. Golec-Biernat and 
J. Kwiecinski, Phys.\ Rev.\ Lett. {\bf 86} (2001) 596.


\bibitem{ActPol}
R.~Venugopalan,
%``Classical methods in DIS and nuclear scattering at small x,''
Acta Phys.\ Polon.\ B {\bf 30}, 3731 (1999).


\bibitem{KMW}
A. Kovner, L. McLerran and H. Weigert,
 Phys. Rev {\bf D52} 3809 (1995); {\bf D52} 6231 (1995).
 Y. V. Kovchegov and D. H. Rischke,  Phys. Rev.  {\bf C56} (1997) 1084;
 M. Gyulassy and L. McLerran, Phys. Rev.  {\bf C56} (1997) 2219.

\bibitem{KN}D. Kharzeev and M. Nardi, Phys. Lett. {\bf B507} (2001), 121; 
D. Kharzeev and E. Levin, nucl-th/0108006; J. Schaffner-Bielich, D. Kharzeev,  
L. McLerran, and R. Venugopalan, nucl-th/0108048.

\bibitem{AR99}
A. Krasnitz and R. Venugopalan, hep-ph/9706329, hep-ph/9808332;
Nucl. Phys. {\bf B557} 237 (1999); 
Phys.\ Rev.\ Lett. {\bf 84} (2000) 4309.

\bibitem{AR01}
A. Krasnitz and R. Venugopalan, Phys. Rev. Lett. {\bf 86} (2001) 1717.

\bibitem{AYR01}
A. Krasnitz, Y. Nara and R. Venugopalan,
 Phys.\ Rev.\ Lett.\ {\bf 87}, 192302 (2001).  

\bibitem{comment1}Our computations are performed for an 
SU(2) gauge theory. 
We expect that $v_2$, since it is a ratio of components of the stress-
energy tensor, will likely be independent of the number of colors.

\bibitem{Lam:1999}
C.~S.~Lam and G.~Mahlon,
Phys.\ Rev.\ D {\bf 61}, 014005 (2000); {\it ibid.}, 
Phys.\ Rev.\ D {\bf 62}, 114023 (2000).

\bibitem{comment3}For the case of classical 
fields, $\alpha$ provides a lower 
bound on $v_2$.

\bibitem{AK97}
J.~Ambj{\o}rn and A.~Krasnitz,
%``Improved determination of the classical sphaleron transition rate,''
Nucl.\ Phys.\ B {\bf 506}, 387 (1997).
%[arXiv:hep-ph/9705380].
%%CITATION = HEP-PH 9705380;%%


\bibitem{comment4}The band will likely be lower for larger values of 
$\Lambda_{s0}R$.

\bibitem{TV}
D.~Teaney and R.~Venugopalan,
%``Classical computation of elliptic flow at large transverse momentum,''
arXiv:hep-ph/0203208.


\bibitem{Voloshin:2002ii}
S.~A.~Voloshin,
%``Anisotropic flow from AGS to RHIC,''
arXiv:nucl-th/0202072.

\bibitem{BMSS}
R. Baier, A. H. Mueller, D. Schiff and D. T. Son, Phys.\ Lett.\ 
{\bf B502} 51 (2001); {\it ibid.}, arXiv:hep-ph/0204211.

\bibitem{JPED}J.~P.~Blaizot and E.~Iancu,
%``The quark-gluon plasma: Collective dynamics and hard thermal loops,''
Phys.\ Rept.\  {\bf 359}, 355 (2002).


\bibitem{Nicola}N.~Borghini, P.~M.~Dinh and J.~Y.~Ollitrault,
%``Flow analysis from multiparticle azimuthal correlations,''
Phys.\ Rev.\ C {\bf 64}, 054901 (2001); Phys.\ Rev.\ C {\bf 63}, 054906 (2001).

\bibitem{KT}Y.~V.~Kovchegov and K.~L.~Tuchin,
arXiv:hep-ph/0203213.

\end{thebibliography}
\end{document}